\def\funp{{I\!\!P}}
\def\xp{x_{{I\!\!P}}}

\def\ovbeta{{\beta}^\prime}

\def\gapprox{\lower .7ex\hbox{$\;\stackrel{\textstyle >}{\sim}\;$}}
\def\lapprox{\lower .7ex\hbox{$\;\stackrel{\textstyle <}{\sim}\;$}}

\newcommand{\be}{\begin{equation}}
\newcommand{\ee}{\end{equation}}
\newcommand{\beeq}{\begin{eqnarray}}
\newcommand{\eeeq}{\end{eqnarray}}

\documentstyle[11pt,epsfig,a4]{article}
\newlength{\dinwidth}
\newlength{\dinmargin}
\begin{document}
\titlepage
\begin{flushright}
DESY 00--180\\
February 2001
\end{flushright}

\vspace*{1in}
\begin{center}
{\Large \bf Diffractive parton distributions from the saturation
model}\\
\vspace*{0.5in}
K. \ Golec-Biernat$^{(a,b)}$
and M. W\"usthoff$^{(c)}$ \\
\vspace*{0.5cm}
{\it $^{(a)}$II.\ Institut f\"ur Theoretische Physik, 
    Universit\"at Hamburg,\\ Luruper Chaussee 149, 
    D-22761 Hamburg, Germany\\
$^{(b)}$H. Niewodnicza\'nski Institute of Nuclear Physics,
   Radzikowskiego 152,\\ 31-342 Krak\'ow, Poland \\
$^{(c)}$Department of Physics, University of Durham,\\ Durham DH1 3LE,
   United Kingdom} \\
\end{center}
\vspace*{2cm}

\vskip1cm 
\begin{abstract}
We review diffractive deep inelastic scattering (DIS) in the light of the
collinear factorization  theorem. This theorem  allows to define 
diffractive parton distributions in the leading twist approach.
Due to its selective final states, diffractive DIS 
offers interesting insight into the form of the diffractive parton
distributions which we explore with the help of the saturation model. 
We find Regge-like factorization with the correct energy dependence
measured at HERA.
A remarkable feature of diffractive DIS is the dominance of the twist-4
contribution for small diffractive masses. We quantify this effect and make a
comparison with the data.

\end{abstract}

\newpage
\section{Introduction}

A significant fraction (around $10\%$) of deep
inelastic scattering (DIS) events observed at HERA at small-$x$ are
diffractive events \cite{H197,ZEUS99}. The proton in these events escapes
almost unscattered down the beam pipe,  losing only a small fraction $\xp$ of
its initial momentum.  The slightly scattered proton is well separated in
rapidity  from the rest of the scattered system, forming a large rapidity gap,
the characteristic feature of diffractive DIS. The ratio of diffractive to all
DIS events is to a good approximation constant as a function of Bjorken-$x$
and $Q^2$.  Thus in a first approximation, DIS diffraction is a leading
twist effect with logarithmic scaling violation. For recent reviews
on diffractive DIS see \cite{REVWU,REVHEB,REVABR}.

Historically, the first description of diffractive DIS was provided
in terms of  the Ingelman-Schlein (IS) model \cite{IS85}. 
The model is based on Regge theory in which diffractive processes are due to
the exchange of a soft pomeron. The novelty of the IS model lies in the
assumption that the pomeron has a partonic structure as do real hadrons.
The diffractive structure function 
 factorizes into a ``pomeron flux'' and
a pomeron structure function. The latter function is written in terms of
the pomeron parton distributions which obey the standard DGLAP evolution
equations. The initial conditions for these equations are 
determined from a fit to the data  
\cite{H197,ZEUS99,ROY00} or using the phenomenology
of soft  hadronic reactions \cite{GK95}.   Despite conceptual difficulties (the
pomeron is not a particle) this idea turned out to be
very fruitful in the description of the data.

The IS model brings the issue of {\it collinear factorization} into 
the leading twist description of DIS diffraction. 
By  this we mean the consistent factorization 
of the diffractive cross section into a convolution of {\it diffractive parton 
distributions} which satisfy the DGLAP evolution equations and computed in
perturbative QCD hard cross sections, in full analogy to inclusive DIS
\cite{BERSOP94,TREVE94,BERSOP96}.
Collinear factorization has been rigorously  
proven by Collins for diffractive DIS \cite{COL98}. Factorization, however, 
fails for hard processes in diffractive hadron-hadron scattering \cite{CFS93}. 
The IS model assumes collinear factorization, imposing 
an additional assumption  on the $\xp$-dependence of the
diffractive parton distributions (called {\it Regge factorization}).
In the general framework of  collinear factorization the diffractive final
state is treated fully inclusive, in particular, 
the mechanism leading to diffraction  is not  elucidated.
The IS model is an attempt to provide such a mechanism.

The detailed description of diffractive processes in DIS,
starting from perturbative QCD, is achieved by modelling the diffractive
final state as well as the interaction with the proton.
Such an analysis goes beyond the leading twist description.
In a first approximation, the diffractive system (separated by the 
rapidity gap from the proton) is formed by a quark-antiquark ($q\bar{q}$) pair
in the color singlet state \cite{NIKO1,MARK1}.
A higher order contribution is represented by the 
$q\bar{q}$ pair with an additional gluon $g$ emitted
\cite{MARK2,NIKO2,BEKW,MARK2a}. 
In the simplest case, the  colorless exchange
responsible for the rapidity gap is modelled by the exchange of two gluons
coupled to the proton with some formfactor \cite{RYSKIN,BARWU} or to a
heavy onium which serves as a model of the proton \cite{HAUT98}.
Higher order corrections are included by the 
the BFKL summation of gluon ladders \cite{MARK3} or using
the color dipole approach \cite{BIALAS}. 
The diffractive processes in DIS
has also been  studied within the semiclassical approach in
\cite{HEB1}. In particular, the relation between
the physical pictures in different frames   was elucidated in
\cite{HEB0}. Let us add that a simple physical interpretation of diffractive
scattering emerges in the proton rest frame, where the formation of
the diffractive system is stretched in time.
Far upstream the beam pipe the virtual photon decays
into a virtual diffractive system which then elastically scatters 
on the proton (without the color exchange), picking up energy to form a real
state.

The  immediate problem faced in such a modelling is the strong sensitivity to 
nonperturbative effects due to the dominance of aligned jet configuration
(discussed in Section~\ref{sec:dpdsm}). Thus, we need a description of
the interactions in the soft regime. The  saturation model \cite{GBW1},
which has already been very successful in describing
inclusive and diffractive DIS data \cite{GBW2}, provides such a
description (see also \cite{JEFF,GENYA,MCDERMOTT} for related
analyses).  Recent
theoretical studies \cite{KOVCHEGOV,KOVCHEGOV1,WEIGERT} justify the 
assumed analytic form of this  model.

In this paper we are going to address questions related to collinear
factorization in the above approach to DIS diffraction.
How to find the  diffractive parton
distributions? What is their form? Do they support
Regge factorization? Is the leading twist contribution sufficient in the
description of DIS diffraction? How important are higher twist contributions?
Answers to all these questions will be found by modeling 
the diffractive system with the help of perturbative QCD. The
interaction with the proton is 
described by the saturation model \cite{GBW1}. From this perspective, we
critically  reexamine the assumptions made in the Ingelman-Schlein model.

Summarizing our results, we find the diffractive quark and 
gluon distributions which serve as the initial
conditions for the  DGLAP evolution equations. Due to a specific form of
the saturation model, the Regge type factorization in $\xp$ is found, 
although Regge theory has not been applied.
Moreover, the correct energy dependence of
diffractive DIS measured at HERA is obtained.
We also perform the numerical analysis for the comparison
with the diffractive data.
As expected, see \cite{BEKW}, the twist-4 contribution from the $q\bar{q}$-pair
produced by longitudinally polarized photons plays a crucial role in the
region of small diffractive mass ($M\ll Q$). The leading twist description
with the DGLAP evolution is insufficient in this regime and the  twist-4
component  (suppressed by $1/Q^2$)
accounts for the difference between  the leading twist
contribution and the data. Thus, there is no need in our
analysis for a singular  (or strongly concentrated at $\beta\approx 1$)
gluon distribution as in the pure leading twist description
\cite{H197}.  The universality of the energy dependence, assumed in the IS
model, is broken by the twist-4 contribution 
which has a steeper $\xp$-dependence than the leading twist contribution. 
The first indication of this effect seems to be observed in the data
\cite{ZEUS99}.

The paper is organized as follows. 
In Section \ref{sec:dpd} we give a general introduction
to diffractive parton distributions. In  Section \ref{sec:dpdsm}
we present the
initial distribution which we extracted from the perturbative QCD approach.
In Section \ref{sec:regge} we make a comment on Regge factorization in the 
context of the saturation model.  
Numerical results are presented in Section \ref{sec:data}
and we finally conclude with a brief summary in Section {\ref{sec:conc}.

\section{Diffractive parton distributions}
\label{sec:dpd}

There are two dimensionful variables, the mass of the diffractive system
$M^2$, and  the momentum transfer $t\;=\; (p-p^\prime)^2$, which characterize
diffractive DIS (see Fig.~\ref{fig:d1}). They come in addition
to $Q^2$ and $W^2$ which are well known from inclusive DIS.

The mass and energy related variables, $M^2$ and $W^2$, are usually rewritten
in terms of the following dimensionless variables:
\be
\label{xp}
\xp \;=\; \frac{Q^2+M^2-t}{Q^2+W^2}\,,  
\ee
\noindent  which describes the fraction of the incident momentum 
lost by the proton or carried by the pomeron, and 
\be
\label{beta}
\beta \;=\; \frac{Q^2}{Q^2+M^2-t}
\ee
\noindent which is the Bjorken variable normalized to the pomeron momentum. 
The true Bjorken variable $x$ connects the two variables
\be
\label{x}
x\;=\;\frac{Q^2}{Q^2+W^2}\;=\;\xp\, \beta\,.
\ee
\noindent In the following  we neglect $t$ in the definition of $\xp$ and
$\beta$ since usually $|t|\ll Q^2,M^2$.
\begin{figure}[t]
  \vspace*{-0.5cm}
     \centerline{
         \epsfig{figure=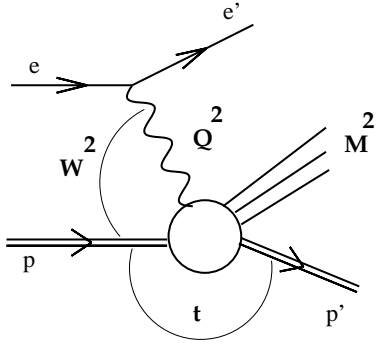,width=5cm}
           }
\vspace*{0.0cm}
\caption{\it Kinematic invariants in DIS diffraction.
\label{fig:d1}}
\end{figure}

The {\it diffractive structure functions} depend on four invariant
variables  $(x, Q^2, \xp, t)$, and their definition follows 
the inclusive counterparts
by relating to the diffractive DIS cross section 
\be
\label{difsf}
\frac{d^4\sigma^D}{dx\, dQ^2\, d\xp\, dt}
\,=\,
\frac{2\pi \alpha^2_{em}}{x\, Q^4}               
\left\{
\left[1+(1-y)^2\right]\;\frac{d\,F_2^{D}}{d\xp\, dt}
\;-\;{y^2}\;\frac{d\,F_L^{D}}{d\xp\, dt}
\right\}\,.
\ee
For simplicity, we introduce the following notation
\beeq 
F_{2,L}^{D(4)}(x, Q^2, \xp, t)\, \equiv \, 
\frac{d\,F_{2,L}^{D}}{d\xp\,dt}(x, Q^2,\xp, t)\,.
\eeeq      
As usual,    
$F_2^{D(4)}\; =\;   F_T^{D(4)}+F_L^{D(4)}$, 
where $T$ and $L$ refer to the polarization of the virtual photon,
transverse and longitudinal, respectively. 
The structure function $F^{D(4)}$
has the dimension $\mbox{\rm GeV}^{-2}$ because of the 
differential $dt$ in the cross section (\ref{difsf}).  The 
dimensionless structure function $F^{D(3)}$ is defined 
by integrating $F^{D(4)}$ over $t$,
\be
\label{difsf3}
F_{T,L}^{D(3)}(x, Q^2, \xp)\,=\,\int dt\;
F_{T,L}^{D(4)}(x,Q^2, \xp, t)\,.
\ee
$F^{D(3)}$ is measured when the final state 
proton transverse momentum is not detected.

The {\it diffractive parton distributions} ${\cal{F}}^D_{a/p}$
 are introduced according to the
{\it collinear factorization} formula \cite{BERSOP94},
\be
\label{eq:dpd}
F_{2}^{D(4)}(x, Q^2, \xp, t)\,=\,
\sum_a\,\int_0^{\xp} d\xi\;
{\cal{F}}^D_{a/p}(\xi, \mu^2,\xp, t)\;
{\cal{C}}_{a}(x/\xi,Q^2/\mu^2)\,,
\ee
with $a=q,g$ denoting a quark or gluon distribution in the proton,
respectively. In the infinite momentum  frame  the diffractive 
parton distributions describe the probability to find a parton  with the
fraction $\xi$ of the proton momentum,  provided the proton stays intact and
loses only a small fraction $\xp$ of its original momentum. ${\cal{C}}_{a}$
are the  coefficient functions   describing hard scattering
of the virtual photon on a parton $a$. 
They are identical to the   coefficient functions known from inclusive DIS,
\be
\label{eq:hcs}
{\cal{C}}_{a}(x/\xi,Q^2/\mu^2)\,=\,e_a^2\,\delta(1-x/\xi)\,+\,
{\cal{O}}(\alpha_{s})\,.
\ee

Formula (\ref{eq:dpd}) is the analogue of the inclusive leading twist
description for inclusive DIS. The inclusive structure function $F_2$ is factorized 
in a similar way into computed in pQCD coefficient functions
and nonperturbative parton distributions. 
The scale $\mu^2$ is the factorization/renormalization scale. Since
the l.h.s of  eq.~(\ref{eq:dpd}) does not depend on this scale, 
i.e. $d F_2^{D(4)}/d\mu^2=0$, we find the
renormalization group equations (evolution equations) for the diffractive
parton distribution
\be
\label{eq:ap}
\mu^2\,\frac{d}{d\mu^2}\; 
{\cal{F}}^D_{a/p}(\xi, \mu^2,\xp, t)
\,=\,
\sum_b \int_{\xi}^{\xp} \frac{dz}{z}\,P_{a/b}(\xi/z,\alpha_s(\mu^2))\;
{\cal{F}}^D_{b/p}(z, \mu^2,\xp, t)\,,
\ee
where $P_{a/b}$ are the standard Altarelli-Parisi splitting functions
in leading or next-to-leading logarithmic approximation. Since the
scale $\mu$ is arbitrary, we can choose $\mu=Q\gg\Lambda_{QCD}$. With this
scale the evolution equations are usually presented.

The integration in (\ref{eq:dpd}) and (\ref{eq:ap})  is only done up to the
fraction $\xp$ of the proton momentum, since the active parton
cannot carry more than this fraction of momentum.
The proton remnants carry the remaining fraction
$(1-\xp)$. If we refer the longitudinal momenta of the partons to 
$\xp p$ instead of the proton total momentum $p$, the structure functions
and parton distributions become functions of $\beta=x/\xp$ or
${\beta}^\prime=\xi/\xp$. With this notation, we rewrite (\ref{eq:dpd}) and
(\ref{eq:ap}) in the following form:    
\be
\label{eq:dpd1}
F_{2}^{D(4)}(\beta, Q^2, \xp, t)\,=\,
\sum_a\,\int_0^{1} d{\beta}^\prime\;
\xp{\cal{F}}^D_{a/p}({\beta}^\prime, \mu^2,\xp, t)\;
{\cal{C}}_a(\beta/{\beta}^\prime,Q^2/\mu^2)\,
\ee
and
\be
\label{eq:ap1}
\mu^2\,\frac{d}{d\mu^2}\; 
{\cal{F}}^D_{a/p}({\beta}, \mu^2,\xp, t)
\,=\,
\sum_b \int_{{\beta}}^1
\frac{dz}{z}\,P_{a/b}({\beta}/z,\alpha_s(\mu^2))\;
{\cal{F}}^D_{b/p}(z,\mu^2,\xp, t)\,.
\ee
Thus, we obtain a description similar to inclusive DIS but modified
by the additional variables $\xp$ and $t$. Moreover, the Bjorken variable $x$
is replaced by its diffractive analogue $\beta$,
eq.~(\ref{beta}). Notice that $\xp$ and $t$ play the role of parameters
of the evolution equations and  does not affect
the evolution. According to the factorization theorem
the evolution equations (\ref{eq:ap1}) are applicable to all orders
in perturbation theory.

In the lowest order approximation for the coefficient functions
(\ref{eq:hcs}), we find for the diffractive structure function
\be
\label{eq:llsf}
F_{2}^{D(4)}(\beta, Q^2, \xp, t)\,=\,\sum_{a=q,\bar{q}}\,e_a^2\;
\beta\;
\xp{\cal{F}}^D_{a/p}({\beta}, Q^2,\xp, t)\,,
\ee
where the sum over the quark flavours is performed.

The collinear factorization formula (\ref{eq:dpd1})
holds to all orders in $\alpha_s$ for diffractive DIS
\cite{COL98}. However, this is no longer true in
hadron--hadron hard diffractive scattering \cite{REVWU,CFS93}, where collinear
factorization  fails due  to final state soft interactions. Thus, unlike 
inclusive scattering,  the diffractive parton distributions are no universal
quantities. The can safely be used, however, to describe hard
diffractive processes involving leptons.
A systematic approach to diffractive parton distributions, based
on quark and gluon operators, is given in \cite{BERSOP96,HAUT98}.

Until now, our discussion has been quite general, in particular
we have not referred to the pomeron. In the 
Ingelman-Schlein (IS) model \cite{IS85}, diffraction is described with the
help of the concept of the soft pomeron exchange. In addition, it is assumed
that the pomeron has a hard structure. In DIS diffraction, this structure is
resolved by the virtual photon, as in the standard DIS processes.
Following the results of Regge theory, the IS model 
is based on the assumption of {\it Regge factorization}. In the context of the
diffractive parton distributions it means that the following factorization
holds \cite{BERSOP94,BERSOP96}
\be
\label{eq:is}
\xp{\cal{F}}^D_{a/p}({\beta}, Q^2,\xp, t)\,=\,
f(\xp,t)\ f_{a/\funp}({\beta},Q^2)\,,
\ee
where the ``pomeron flux'' $f(\xp,t)$ is given by
\be
f(\xp,t)\,=\,
\frac{B^2(t)}{8\pi^2}\;\xp^{1-2\alpha_\funp(t)}\,. 
\ee
Thus, the variables $(\xp,t)$,
related to the loosely scattered proton, are factorized from 
the variables characterizing the diffractive system $(\beta,Q^2)$.
$B(t)$ is the Dirac electromagnetic form factor \cite{DL84}, 
$\alpha_\funp(t)=1.1+0.25~\mbox{\rm GeV}^{-2}\cdot t$ is the soft pomeron
trajectory \cite{DL92} and 
the normalization  of  $f(\xp,t)$ follows the 
convention of \cite{DL88}. The function $f_{a/\funp}(\beta,Q^2)$ 
in eq.~(\ref{eq:is}) describes the
hard structure in DIS diffraction, and  is interpreted
as the pomeron parton distribution.
Now, the diffractive
structure function (\ref{eq:llsf}) becomes
\be
\label{eq:isfd}
F_{2}^{D(4)}(\beta, Q^2, \xp, t)\,=\,f(\xp,t)
\;
\sum_{a=q,\overline{q}}\,e_a^2\,\beta\,f_{a/\funp}(\beta,Q^2)\,,
\ee
where the summation over quarks and antiquarks is performed.
The  $Q^2$-evolution of $f_{a/\funp}(\beta,Q^2)$ is given by the 
DGLAP equations (\ref{eq:ap1}). The $t$-dependence in the pomeron
parton distributions is neglected. 

The pomeron parton distributions are determined  as the parton
distributions of  real hadrons. Some functional form with several
parameters  is assumed at an initial scale
 and then the  parameters are found from a fit to data
\cite{H197,ZEUS99,ROY00} using the DGLAP evolution equations.
An alternative  approach for the determination of the initial
distributions makes  use of the phenomenology of soft hadronic reactions
\cite{GK95}.

Ingelman and Schlein have conceived of their approach primarily to describe
diffractive hard scattering at hadron colliders. This includes the 
concept of Regge factorization as well as pomeron parton distributions.
Unfortunately, collinear factorization was proven to be wrong in the
case of hard diffraction at hadron colliders. Although it holds for diffractive DIS,
this does not imply Regge factorization nor the existence of pomeron
parton distributions. Our approach discussed below
does not make use of either Regge factorization or pomeron parton
distributions. It does, however, result in diffractive parton distributions
and $\xp$-factorization.


\section{Diffractive parton distributions and the saturation model}
\label{sec:dpdsm}

In  ref. \cite{GBW1} an analytic form for the dipole cross section 
$\hat{\sigma}(x,r)$ was suggested, based
on the idea of saturation \cite{GLR}, which allows to describe the
proton structure function $F_2$ at small $x$, 
\be
F_{2}(x,Q^2) = \frac{Q^2}{4\pi^2\alpha_{em}}
\int d^2{\bf{r}}\, dz\, \left(|\Psi_T(r,z,Q^2)|^2+|\Psi_L(r,z,Q^2)|^2\right)
\, \hat{\sigma}(x,r),
\ee
where $\Psi_{T,L}$ is the $q\bar{q}$ dipole
wave function for transverse $(T)$ and longitudinal $(L)$ photons.
$r$ is the dipole transverse size and $z$ is a fraction of the virtual photon
momentum carried by a quark (antiquark).
A distinctive feature of the  dipole cross section is its scaling
form, i.e. 
\be
\label{eq:dipcs}
\hat{\sigma}(x,r)\;=\;{\sigma_0}\;g(r/R_0(x))
\;=\;{\sigma_0}\,\left\{1-\exp(-r^2/R_0^2(x)\right\}
\,,
\ee
where the function $R_0(x)=1/\mbox{\rm GeV}\,(x/x_0)^{\lambda/2}$ monotonically
vanishes when  $x\rightarrow 0$, and $\sigma_0$ is an overall
normalization. The three parameters of $\sigma_0=23~\mbox{\rm mb}$,
$x_0=3\cdot 10^{-4}$ and $\lambda=0.29$ were found form a fit   to inclusive
DIS  data at small-$x$. 

The postulated form 
allows the transition to saturation where $\hat\sigma=\sigma_0$ for
large dipole sizes, while for small $r$ color transparency
is assumed, $\hat\sigma\sim r^2$. 
The fact that the dipole cross section  depends on $r$ and $x$
through the dimensionless  ratio $r/R_0(x)$ leads to the prediction  
of a new scaling property for small-$x$ data \cite{SGK00}.

The presented model successfully describes $F_2$ at small $x$, including
the region of small $Q^2$ values.
With regard to diffractive processes in DIS, $\hat{\sigma}$ leads to 
the same dependence on $x$ and $Q^2$
of diffractive cross section as for  inclusive DIS, 
and gives a good description of the data \cite{GBW2}. Thus, 
the constant ratio between the diffractive and inclusive cross sections
finds a natural explanation in this model.

\begin{figure}[t]
  \vspace*{-0.5cm}
     \centerline{
         \epsfig{figure=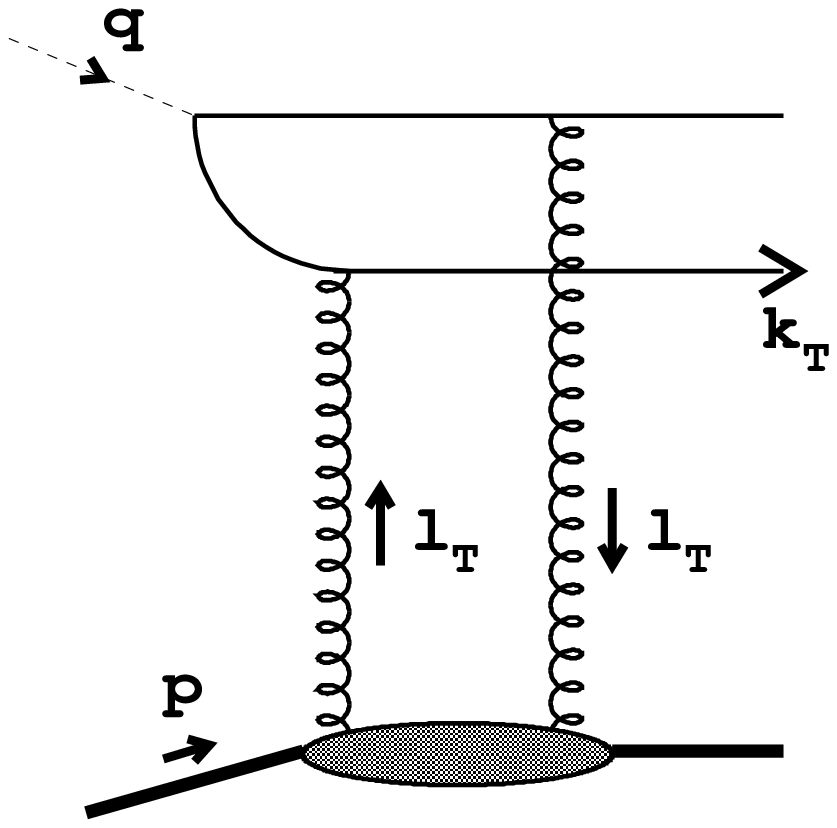,width=6cm} 
         \epsfig{figure=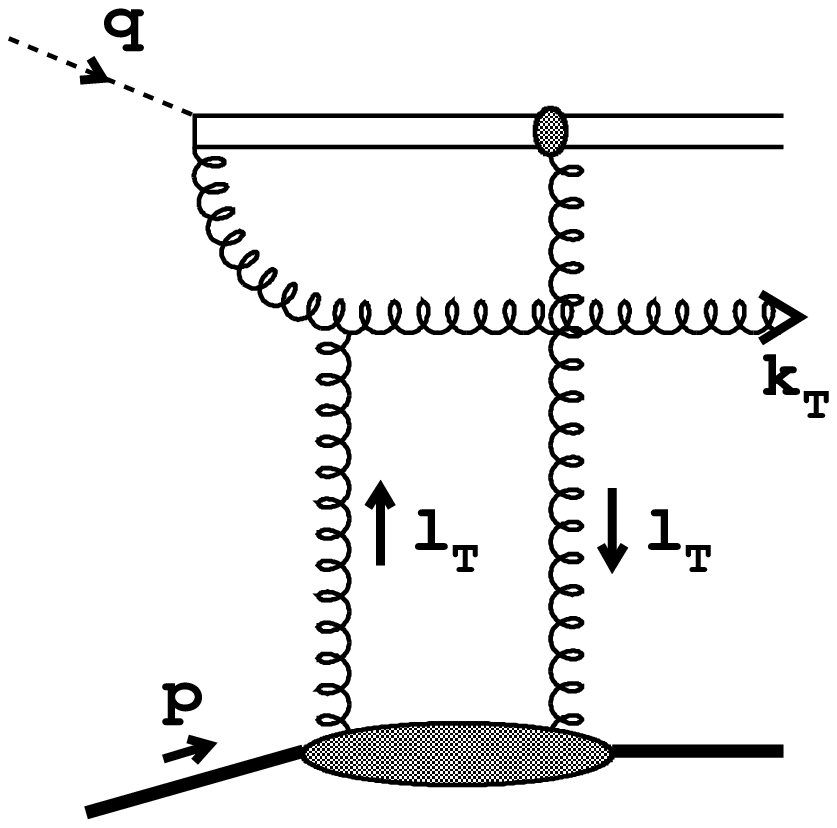,width=6cm}
           }
\vspace*{0.0cm}
\caption{\it The diffractive $q\bar{q}$ and $q\bar{q}g$ contributions to
$F_2^{D(3)}$. 
\label{fig:diag1}}
\end{figure}

Following the idea of the analysis \cite{BEKW}, 
the diffractive structure function $F_2^{D(3)}$ is the sum
of the three contributions shown in Fig.~\ref{fig:diag1}, the $q\bar{q}$ 
production from transverse and longitudinal photons, and
the $q\bar{q}g$ production,
\be
\label{eq:f2dtot}
F_2^{D(3)}(\beta,Q^2,\xp)\,=\,
F^T_{q\bar{q}}\,+\,F^L_{q\bar{q}}\,+\,F^T_{q\bar{q}g}\,,
\ee
where $T$ and $L$ refer to the polarization of
the virtual photon. For the $q\bar{q}g$ contribution only the transverse
polarization is considered, since the longitudinal counterpart has no leading
logarithm in $Q^2$. In this approach, the diffractive $q\bar{q}$ and
$q\bar{q}g$   systems  interact with the proton
like in the two gluon exchange model. 
The coupling of the two gluons to the proton is effectively
described the dipole cross section, determined from the analysis of inclusive
DIS.

The computation of the diffractive structure functions in this case
was presented in \cite{GBW2}. Here we quote only the
final results.
The transverse $q\bar{q}$ part is given by 
\beeq
\label{ftqq}
F_{q\bar{q}}^{T}
\,=\,
\frac{3}{64\pi^4 B_D\,\xp}\;\sum_f e_f^2 \;
\frac{\beta^2}{(1-\beta)^3}\;
\int_0^{\frac{Q^2(1-\beta)}{4\,\beta}} dk^2\;
\frac{\displaystyle 1-\frac{2\beta}{1-\beta}\frac{k^2}{Q^2}}
{\displaystyle \sqrt{1-\frac{4\beta}{1-\beta}\frac{k^2}{Q^2}}}\;\;
\phi_1^2(k,\beta,\xp)\,, 
\eeeq
and the longitudinal $q\bar{q}$ contribution takes the form
\be
\label{flqq}
F_{q\bar{q}}^{L}\,=\,
\frac{3}{16\pi^4 B_D\,\xp}\;\sum_f e_f^2 \;
\frac{\beta^3}{(1-\beta)^4}\;
\int_0^{\frac{Q^2(1-\beta)}{4\,\beta}} dk^2\; 
\frac{\displaystyle {k^2}/{Q^2}}
{\displaystyle \sqrt{1-\frac{4\beta}{1-\beta}\frac{k^2}{Q^2}}}\;\;
\phi_0^2(k,\beta,\xp)\,,
\ee
where the ``impact factors'' $\phi_{i=0,1}$ read
\be
\label{eq:imfac}
\phi_{i}(k,\beta,\xp)
\;=\;k^2\,
\int_0^\infty dr\, r\, K_{i}\!\left(\sqrt{\frac{\beta}{1-\beta}}kr\!\right)\,
J_{i}(kr)\;  \hat{\sigma}(\xp,r)\;,
\ee
and $K_{i}$ and $J_{i}$ are the Bessel functions. The quoted formulae
correspond to eqs.~(32) and (33) in \cite{GBW2}, respectively, with
the angular integration already done, see also the second paper of \cite{JEFF}
for a similar result. $B_D$ is the diffractive slope, present due to the
integration over $t$ of  $F^{D(4)}$ with the assumed $t$-dependence:
$F^{D(4)}=\exp(B_D t)\,F^{D(3)}$. Its value is taken from the HERA experiments.
The $k$-integration in eqs.~(\ref{ftqq}) and  (\ref{flqq}) is
performed over the transverse momentum of the quarks in the $q\bar{q}$ pair,
$k=|{\bf k}_\perp|$. 
The final result 
depends on the squared ``impact factors'', and thus on the square
of the  dipole cross section $\hat\sigma$.

One should realize 
that the longitudinal contribution is suppressed by a power of
$Q^2$ in comparison to the transverse contribution, i.e.  the {\it longitudinal
structure function} is a {\it higher twist} contribution. Although of higher
twist nature this contribution has some importance as will be seen later.  

The {\it leading
twist} part of the {\it transverse structure function}, which corresponds to the
diffractive $q\bar{q}$ production, can be extracted from (\ref{ftqq}) by
neglecting the factors with powers  of $k^2/Q^2$ under the integral and taking
the upper limit of the integration to infinity. Strictly speaking, energy
conservation is violated in this case, but the  corrections are of higher
twist nature which  in this case will be neglected because of their smallness. 
With the new limit the integral is still finite, and we can write 
the leading twist part of (\ref{ftqq}) as
\be
\label{ftqqlead}
F_{q\bar{q}}^{T(LT)}
\,=\,
\frac{3}{64\pi^4 B_D\,\xp}\;\sum_f e_f^2 \;
\frac{\beta^2}{(1-\beta)^3}\;
\int_0^{\infty} dk^2\;  
\phi_1^2(k,\beta,\xp)\;.
\ee
A detailed analysis based on the dipole representation of the
$q\bar{q}$ wave function shows that the approximation
leading to eq.~(\ref{ftqqlead}) corresponds to the {\it aligned jet
configuration} of the $q\bar{q}$ pair in the proton rest frame. 
The smallness of the factor 
\be
\frac{\beta}{1-\beta}\,\frac{k^2}{Q^2}\,=\,\frac{k^2}{M^2}\,=\,z(1-z)
\ee
which we neglected in eq.~(\ref{ftqq}) ($M$ is the diffractive mass 
and $z,\, (1-z)$ are the longitudinal momentum
fractions of the final state quarks with respect to the photon momentum) 
 means that
one of the quarks takes almost the whole longitudinal 
photon momentum (e.g. $z\approx 1$) while
the other quark forms the remnant with $(1-z)\approx 0$. Similar conclusions on
the leading twist part of diffractive DIS have been drawn in ref.~\cite{HEB0}.

Now, we can determine the diffractive quark distributions according to
eq.~(\ref{eq:llsf}) (integrated over $t$),
\be
\label{eq:dpd2}
F_{q\bar{q}}^{T(LT)}\,=\,2\,
\sum_{f} e_f^2\;\beta\; q^D(\beta,Q^2,\xp)
\ee
Hence the {\it diffractive quark distribution}  is given by
(independent of the quark flavour $f$)
\be
\label{eq:pdq}
q^D(\beta,Q^2,\xp)\,\equiv\,
\xp{\cal{F}}^D_{q/p}(\beta,Q^2,\xp)\,=\,
\frac{3}{128\pi^4 B_D\,\xp}\;
\frac{\beta}{(1-\beta)^3}\;
\int_0^{\infty} dk^2\; 
\phi_1^2(k,\beta,\xp)\,,
\ee
Notice the lack of the $Q^2$-dependence on the r.h.s.
of eq.~(\ref{eq:pdq}). This may be viewed as  a consequence of
not having included ultraviolet divergent corrections which would require a
cutoff. With those corrections the parton distributions become
$\mu^2$-dependent, and evolution would relate the distributions at different
$\mu^2$ values. Still, we may use the found diffractive quark
distributions as the input distributions for the DGLAP evolution equations
at some initial  scale. Of course,   the
choice of the initial scale introduces an  uncertainty for the prediction.

The detailed  discussion of the $q\bar{q}g$ contribution can be found 
in \cite{GBW2} and \cite{MARK2} with the details of the calculations in the Appendix.
The new contribution was computed assuming  strong
ordering in transverse momenta of the gluon and the $q\bar{q}$ pair,
i.e. $k_{\perp g} \ll k_{\perp q} \approx k_{\perp \bar{q}}$. This assumption
allows to treat the $q\bar{q}g$ system as a $g\bar{g}$ dipole in the transverse
configuration space $\bf{r}$, where $\bf{r}$ is the Fourier conjugate
variable to the quark transverse momentum $\bf{k}_\perp$.

The formula for the $q\bar{q}g$ diffractive structure function
which we quote below corresponds to eq.~(39) in \cite{GBW2},
integrated over the azimuth angle in configuration space
\footnote{
A factor 1/2 missing in eq.(39) of ref.~\cite{GBW2} was correctly pointed out
in ref.~\cite{JEFF}. This does not affect the numerical results in
\cite{GBW2}.}.
Thus we have 
\beeq
\label{f3dgl}
F_{q\bar{q}g}^{T}(\beta,Q^2,\xp)
&=&
\frac{81 \beta}{256\pi^4 B_D\,\xp}\;
\sum_f e_f^2\;\;\frac{\alpha_s}{2\pi}\;
\int_\beta^1 \frac{d\ovbeta}{\ovbeta} 
\left[\left(1-\frac{\beta}{\ovbeta}\right)^2
+\left(\frac{\beta}{\ovbeta}\right)^2\right]
\\ \nonumber
\\ \nonumber
&\times&
\frac{\ovbeta}{(1-\ovbeta)^3}\;
\int_0^{Q^2(1-\ovbeta)} dk^2\; \log\left(\frac{Q^2(1-\ovbeta)}{k^2}\right)\;
\phi_2^2(k,\ovbeta,\xp)\;,
\eeeq
\noindent where the impact factor $\phi_2$ is given by
eq.~(\ref{eq:imfac}) with $i=2$.
The variable $\ovbeta$ describes the momentum fraction of the $t$-channel
exchanged gluon with respect to the pomeron momentum $\xp p$. The combination
$k^2/(1-\ovbeta)$ which enters the logarithm is its mean virtuality, and
$k=|{\bf k}_\perp|$ is the transverse momentum of the final state
gluon. The term
in square brackets under the first integral is the Altarelli-Parisi
splitting function for $g\rightarrow qq$, which results from the approximation
that the transverse momentum of the emitted gluon is smaller than
transverse momenta of the quarks.

The diffractive gluon distribution can be found from eq.~(\ref{f3dgl}). In the
calculation of this contribution strong ordering between the gluon and
quark transverse momenta was assumed. In this
approximation the integration over the transverse momentum of the quark loop
gives a logarithmic contribution which has a natural lower cutoff, the
virtuality of the gluon $k^2/(1-\ovbeta)$. At the same time the virtuality of
the gluon should not exceed $Q^2$. This is the origin of the logarithmic term
in (\ref{f3dgl}). Collinear factorization means that we can pull that logarithm
out of the integral over the gluon transverse momenta,
and add to it an arbitrary initial  scale $Q_0^2< Q^2$. Thus we can write 
\be
\label{colfdgl}
F_{q\bar{q}g}^{T}
\,=\,
2\,
\sum_f e_f^2\;\beta\, \left\{
\frac{\alpha_s}{2\pi}\; \log\frac{Q^2}{Q_0^2}\; 
\int_\beta^1 \frac{d\ovbeta}{\ovbeta}\;\frac{1}{2}
\left[\left(1-\frac{\beta}{\ovbeta}\right)^2
+\left(\frac{\beta}{\ovbeta}\right)^2\right]\,
\xp{\cal{F}}_{g/p}^D(\ovbeta,\xp)
\right\}\,,
\ee
\noindent where the {\it diffractive gluon distribution}
is given by 
\be
\label{eq:pdg}
g^D({\beta},\xp)\,\equiv\,
\xp{\cal{F}}_{g/p}^D({\beta},\xp)
\;=\;
\frac{81}{256\pi^4 B_D\,\xp}\;
\frac{{\beta}}{(1-{\beta})^3}\;
\int_0^{\infty} dk^2\; \phi_2^2(k,{\beta},\xp)\,
\ee
and $\phi_2$ is given by eq.~(\ref{eq:imfac}) with $i=2$.
As in the case of the quark distribution (\ref{eq:pdq}),
the found gluon distribution does not depend on $Q^2$, and
serves as the initial
distributions at some fixed scale $Q_0^2$.

The motivation for the above identification of the diffractive 
gluon distributions is the structure in the curly brackets
on the r.h.s of eq.~(\ref{colfdgl}).
It is identical to the structure resulting from the DGLAP evolution 
with one splitting of the gluon into a $q\bar{q}$ pair. 

\begin{figure}
   \vspace*{-1cm}
    \centerline{
     \epsfig{figure=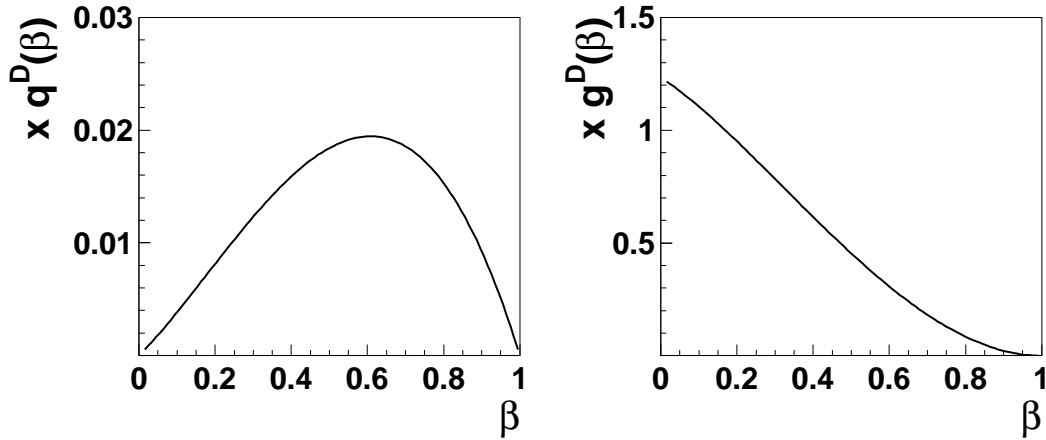,width=15cm}
               }
    \vspace*{-0.5cm}
\caption{Diffractive quark (\ref{eq:pdq})
and gluon (\ref{eq:pdg}) distributions (multiplied by $x=x_\funp\, \beta$)
based on the saturation model as a function of $\beta$ for $\xp=0.0042$ at the
initial scale $Q_0^2$. }
\label{fig:dpd}
\end{figure}

The combined initial parton distributions (\ref{eq:pdq}) and
(\ref{eq:pdg}) ( depicted in Fig.\ref{fig:dpd} ) allow a complete description 
of the leading twist part of diffractive DIS by serving as the initial
conditions for  the DGLAP evolution equations. 
DGLAP evolution means that our diffractive system becomes more complicated
due to additional parton emissions. 

The longitudinal, higher twist contribution requires a separate treatment.
It becomes important for large values of $\beta$,
where the $q\bar{q}$ and the $q\bar{q}g$ production from transverse
photons is negligible \cite{GBW2,MARK2,BEKW}. 
In our present analysis we simply add this contribution to the evolved leading
twist part. The complete expression of the structure function reads
\be
\label{eq:newanal}
F_2^{D(3)}\,=\,F_2^{D(3)(LT)}\,+\,F_{q\bar{q}}^{L}\,.
\ee
where $F_2^{D(3)LT}$ is given by \be
\label{eq:finlt}
F_2^{D(3)(LT)}
\,=\,
2\, \sum_f e_f^2\;\beta\;q^D(\beta,Q^2,\xp)\,,
\ee
with the full  DGLAP evolution. $F_{q\bar{q}}^{L}$ is given by
eq.~(\ref{flqq}).

In the following we present a comparison of the description based on
eq.~(\ref{eq:newanal})
with the diffractive data from HERA. Before doing that we briefly 
discuss $\xp$-factorization.

\section{The issue of  Regge factorization}
\label{sec:regge}

Up to now,  we have not made use of the
particular form (\ref{eq:dipcs}) of the dipole cross section $\hat\sigma$.
We only implicitly assumed that the integrals involving
$\hat\sigma$ are finite. The scaling property, i.e. that
$\hat\sigma$ is a function of the dimensionless ratio  $r/R_0(x)$,  has 
the remarkable consequence that the $\xp$-dependent part of the found diffractive
parton  distributions (DPD) factorizes.

Introducing the dimensionless variables $\hat{k}=k R_0(x)$ and $\hat{r}=r/R_0(x)$
in (\ref{eq:pdq}) and (\ref{eq:pdg}) and assuming $Q_0^2$ to be a fixed scale, 
we find the following factorized form
\beeq
q^D(\beta,\xp) &=& \frac{1}{\xp R_0^2(\xp)}\;f_{q/\funp}(\beta)\,,
\\ \nonumber
\\
g^D(\beta,\xp) &=& \frac{1}{\xp R_0^2(\xp)}\;f_{g/\funp}(\beta)\,.
\eeeq
We have introduced a notation similar to that in (\ref{eq:is}) for the 
$\beta$-dependent factors. This type of factorization looks like Regge factorization
but has nothing to do with Regge theory. It merely results from the scaling
properties of the saturating cross section $\hat\sigma$.
Since the evolution does not affect the
$\xp$-dependence of the DPD, the factorized form will be valid for any scale
$Q^2$.

Now, we can rewrite eq. (\ref{eq:finlt}) as
\be
\label{eq:ltwsm}
F_2^{D(3)(LT)}\,=\,
\frac{1}{\xp R_0^2(\xp)}\,2\,\sum_f e_f^2\;\beta\,f_{q/\funp}(\beta,Q^2)\,
\ee
in which the $\xp$-dependence is factored out. In the saturation model
($R_0(x)\sim x^{\lambda/2}$) the parameter $\lambda=0.29$ was determined from a
fit to inclusive DIS data only \cite{GBW1}. The same value holds for
diffractive interactions, thus we find a definite prediction for
the $\xp$-dependence of the leading twist diffractive structure function
\be
\label{eq:poms1}
F_2^{D(3)(LT)}\,\sim\,
\xp^{-1-\lambda}\,.
\ee

At present, the bulk of diffractive data in DIS support 
the factorized form (\ref{eq:poms1}). They are usually interpreted \cite{H197, ZEUS99} 
in terms of the $t$-averaged  pomeron intercept $\overline{\alpha_\funp}$, i.e.
\be
\label{eq:poms2}
F_{2}^{D(3)}\,\sim\,\xp^{1-2\,\overline{\alpha_\funp}}\,.
\ee
Such a dependence has been introduced in the spirit of the Ingelman-Schlein
model (\ref{eq:isfd}), with the  $t$-integration performed,
$F_{2}^{D(3)}\sim \int dt\,f(\xp,t)\sim \xp^{1-2\,\overline{\alpha_\funp}}$.
Thus, according to (\ref{eq:poms1}) and (\ref{eq:poms2}) we find 
\be
\overline{\alpha_\funp}\,=\,\frac{\lambda}{2}+1\,\approx\,1.15\,,
\ee
which is in remarkable agreement with the values found at HERA,
$\overline{\alpha_\funp}=\alpha_\funp(0)-0.03=1.17$ by H1 \cite{H197} and
$\overline{\alpha_\funp}=1.13$ by ZEUS \cite{ZEUS99}.

Summarizing, the leading twist description extracted from the saturation model
of DIS diffraction leads to the factorization of the $\xp$-dependent part of
the cross section similar to Regge factorization. It correctly predicts 
the value of the 
``effective pomeron intercept''. The $Q^2$-dependence of the diffractive
structure function does not affect the $\xp$-factorization.
This means that the saturation model for the dipole cross section
gives effectively the result which coincides 
with the Regge approach to DIS diffraction, although the physics behind is
completely different. The relative hardness of the intrinsic scale 
$1/R_0(\xp)\sim 1~\mbox{\rm GeV}$  
in the saturation model suggests 
that DIS diffraction is a semihard process rather than a soft process as Regge 
theory would require.

In the presented description, the leading twist structure function
vanishes when $\beta\rightarrow 1$, i.e. for small diffractive mass $M^2\ll
Q^2$. This is not the case for the
higher twist longitudinal contribution $F_{Lq\bar{q}}^D$\,, eq.~(\ref{flqq}),
which dominates in the region of $\beta\approx 1$ \cite{BEKW}.
The expected $\xp$-dependence for this contribution is given by
\be
F_{q\bar{q}}^{L}\,\sim\,
\frac{1}{\xp R^4_0(\xp)}\,\sim\,\xp^{-1-2\lambda}\,,
\ee
which clearly violates the universality of the effective pomeron intercept,
assumed in the Ingelman-Schlein model. The first indication of this
effect is  observed at HERA by measuring the effective pomeron intercept
in different regions of the diffractive mass as a function
of $Q^2$ \cite{ZEUS99}. The intercept seems to be bigger for smaller 
diffractive masses ($\beta\rightarrow 1$), and  the description based on 
perturbative QCD gives a satisfactory explanation \cite{GBW2}.

\section{Comparison with data}
\label{sec:data}

In this section we present a comparison of 
the leading twist part of diffractive DIS
including DGLAP evolution plus the corresponding longitudinal twist-4 component
(eq.~(\ref{eq:newanal})) with the diffractive data from the HERA experiments.

In Fig.~\ref{fig:dpd} we show the quark and gluon diffractive
distributions at some initial scale $Q_0^2$.
This scale, however, is not determined in this approach. Thus, it can be treated
as a phenomenological parameter which may be tuned to obtain the best
description of the DIS diffractive data. 
We tried various options and found that 
$Q_0^2\approx 1.5~\mbox{\rm GeV}^2$ is the best choice for this purpose\footnote{This value has been changed
in comparison to the journal version because of a numerical error in the analysis there. The results of a comparison with the data
are practically the same.}.
We used
the leading logarithmic evolution equations 
with three massless flavours, and
the value of $\Lambda_{QCD}=200~\mbox{MeV}$ in $\alpha_s$.

Fig.~\ref{fig4} shows the results of our studies with data 
from H1 and Fig.~\ref{fig5} data from ZEUS. 
The dashed lines represent the total contribution 
according to eq.~(\ref{eq:newanal}). The pure
leading twist structure function $F_2^{D(3)LT}$ with the leading
DGLAP evolution is shown by the dotted lines. 
The difference (if visible) between the dashed and
dotted lines is the effect of the longitudinal twist-4 component
$F^L_{q\bar{q}}$ added into the leading twist result. As expected, the twist-4
component is significant in the large-$\beta$ domain, see also \cite{BEKW}
for a detailed discussion. Notice the change of the slope in $\xp$ when
the twist-4 component is added. 
The overall agreement between the data and the model (\ref{eq:newanal}) is 
reasonably good, taking into account the fact that the only tuned parameter
is the initial scale for the evolution $Q^2_0$. The parameters of the evolution
equations are standard, and the diffractive slope $B_D=6~\mbox{\rm GeV}^{-2}$
which effects the normalization is taken from the measurements as in the
analysis \cite{GBW2}. The significance of the twist-4
contribution at large $\beta$ is clearly demonstrated. 

Looking at Fig.~\ref{fig4} and
\ref{fig5}, we realize that at large $\beta$ and $Q^2$ the agreement between
the data and the model with evolution starts to deteriorate.
The reason for this
is illustrated in Fig.~\ref{fig6} where we show $F_2^D$ for fixed value of
$\xp=0.0042$ as  a function of $\beta$, for different values of $Q^2$. The
DGLAP evolution depopulates the region of large $\beta$, shifting the parton
distributions towards smaller values  of $\beta$ (compare the dotted lines
showing the initial distributions and the dashed lines showing 
the evolved distributions).
The twist-4 component $F^L_{q\bar{q}}$ 
(the difference between the solid
and the dashed lines) largely compensates for this effect.   Its
significance, however,  diminishes as $Q^2$ rises due to the $1/Q^2$ dependence
of twist-4. 

There are at least two effects which
may enhance $F^L_{q\bar{q}}$ when $Q^2$ rises, thus
accounting for the difference between the data and the discussed description.
In principle, twist-4
should also have logarithmic evolution, in addition to the $1/Q^2$ dependence,
which could push $F^L_{q\bar{q}}$ into the right direction.
Unfortunately, this aspect is beyond the scope of the
present paper.  Another effect which is important at large $\beta$ is
skewedness of the diffractive parton distributions, see \cite{HEBTEU} for a 
recent discussion and references therein. This effect is known to enhance
$F_{q\bar{q}}^{T,L}$ at large $\beta$. Since the enhancement rises with
$Q^2$, the skewedness may account for the discussed difference between the data
and their description in the region of large $\beta$ and $Q^2$.

\section{Conclusions}
\label{sec:conc}

We have reviewed the description of diffractive deep
inelastic scattering in the light of the collinear factorization theorem.
This theorem applies to the leading twist terms of the cross section and
introduces the notion
of diffractive parton distributions. We have extracted a precise analytic form
for these distributions from the approach in which the diffractive state
is formed by the $q\bar{q}$ and $q\bar{q}g$ systems, computed in pQCD.
The convolution 
with a dipole cross section from the saturation model
leads to the $\xp$-factorization of the diffractive structure functions,
similar to Regge factorization,  
which correctly describe the  energy dependence found at HERA by the H1 and
ZEUS collaborations. 
We further evolved
the diffractive parton distributions with the DGLAP evolution equations 
and pointed out the significance of the
twist-4 component at large $\beta$ for the agreement with the data.
The latter was
originally advocated in \cite{BEKW} and originates from
the longitudinal $q{\bar{q}}$ contribution. The twist-4 component
breaks the
universality of the effective $\xp$-dependence, making it stronger at large
$\beta$. This stays in contrast to the assumed universality in the
Ingelman-Schlein model.
Finally, we suggest 
some necessary modifications
of the description at large $\beta$ and $Q^2$ to improve the agreement with
the data in this kinematic region.

\vskip 1cm
\centerline{\large \bf Acknowledgements}

We thank Jochen Bartels and Henri Kowalski for useful discussions and critical
reading of the manuscript.
K.G.-B. thanks  Deutsche Forschungsgemeinschaft
for financial support.
This research has been supported in part by the EU
Fourth Framework Programme  
``Training and Mobility of Researchers''  Network,    
``Quantum Chromodynamics and the Deep Structure of Elementary
Particles'', contract   FMRX-CT98-0194 (DG~12-MIHT) and by the Polish KBN
grant No.  5 P03B 144 20.


\begin{figure}
   \vspace*{-1cm}
    \centerline{
     \epsfig{figure=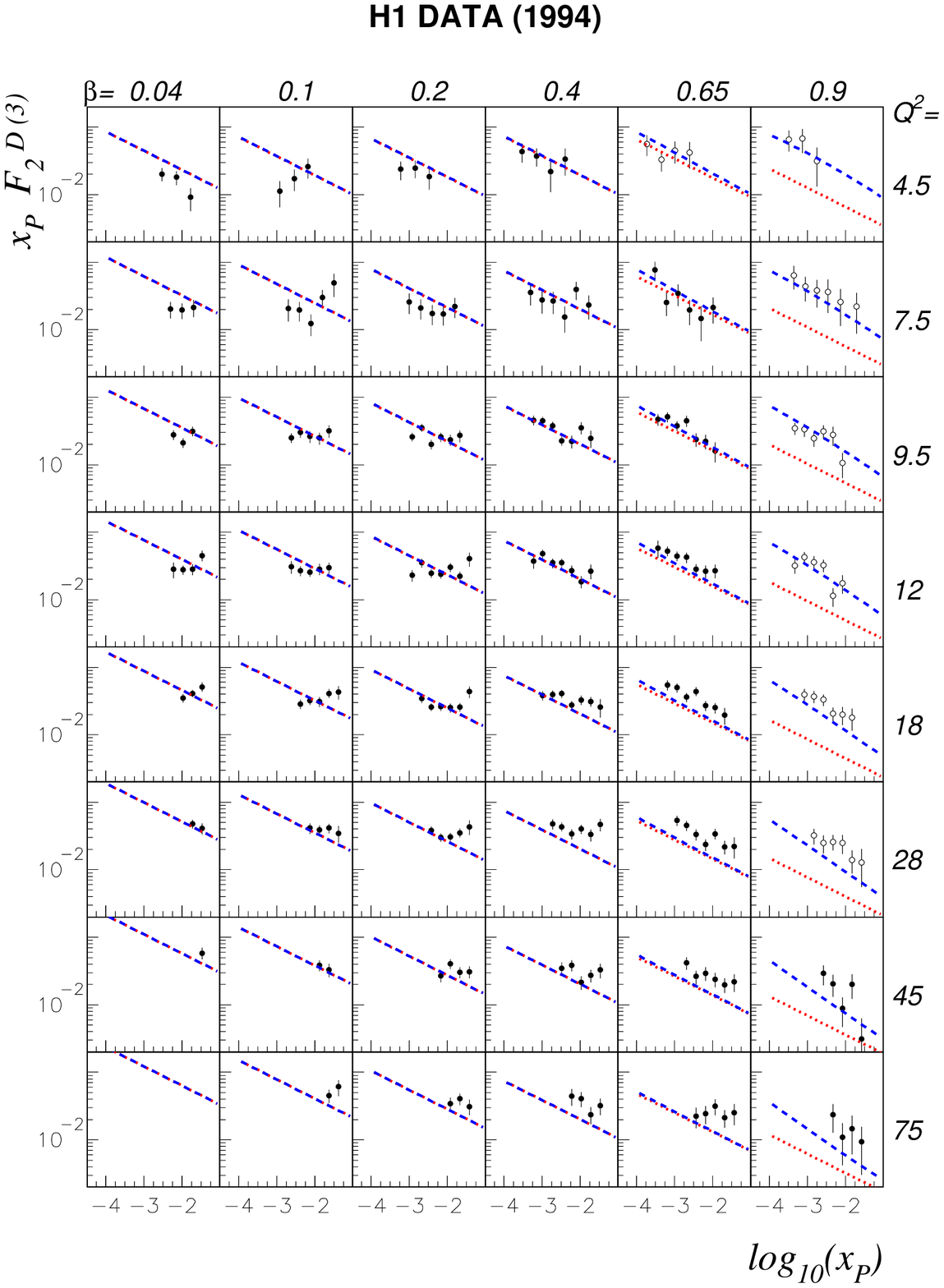,width=15cm}
               }
    \vspace*{0cm}
\caption{The comparison with H1 data \cite{H197}. 
The dashed lines
correspond to the leading twist contribution with the
twist-4 component added, eq.~(\ref{eq:newanal}). The leading twist
contribution is shown by the dotted lines.
}
\label{fig4}
\end{figure}

\begin{figure}

   \vspace*{-1cm}
    \centerline{
     \epsfig{figure=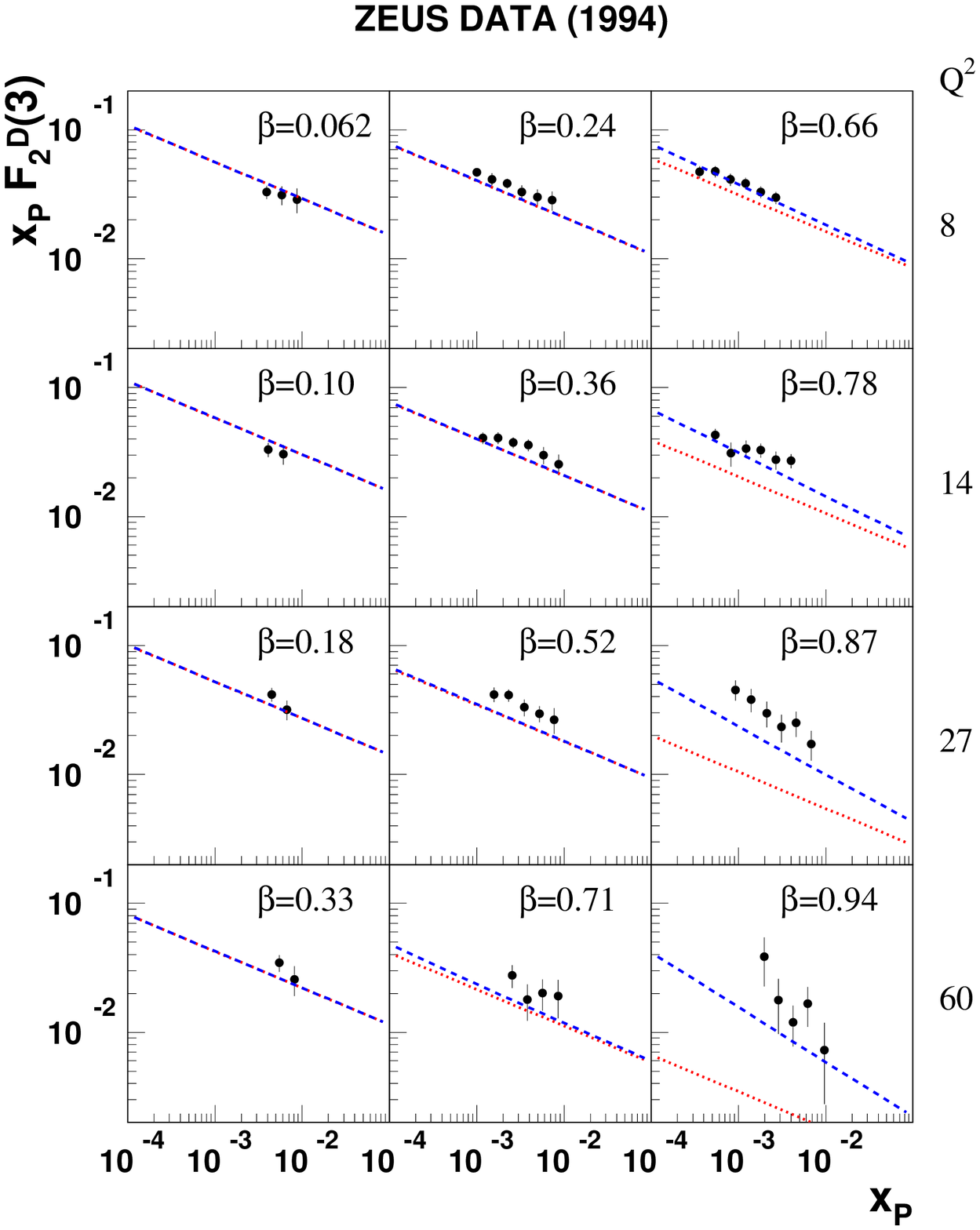,width=15cm}
               }
    \vspace*{0cm}
\caption{The comparison with ZEUS data \cite{ZEUS99}. 
The dashed
lines correspond to the leading twist contribution with
the twist-4 component added, eq.~(\ref{eq:newanal}). The leading
twist contribution is shown by the dotted lines.
}
\label{fig5}
\end{figure}

\begin{figure}
   \vspace*{-1cm}
    \centerline{
     \epsfig{figure=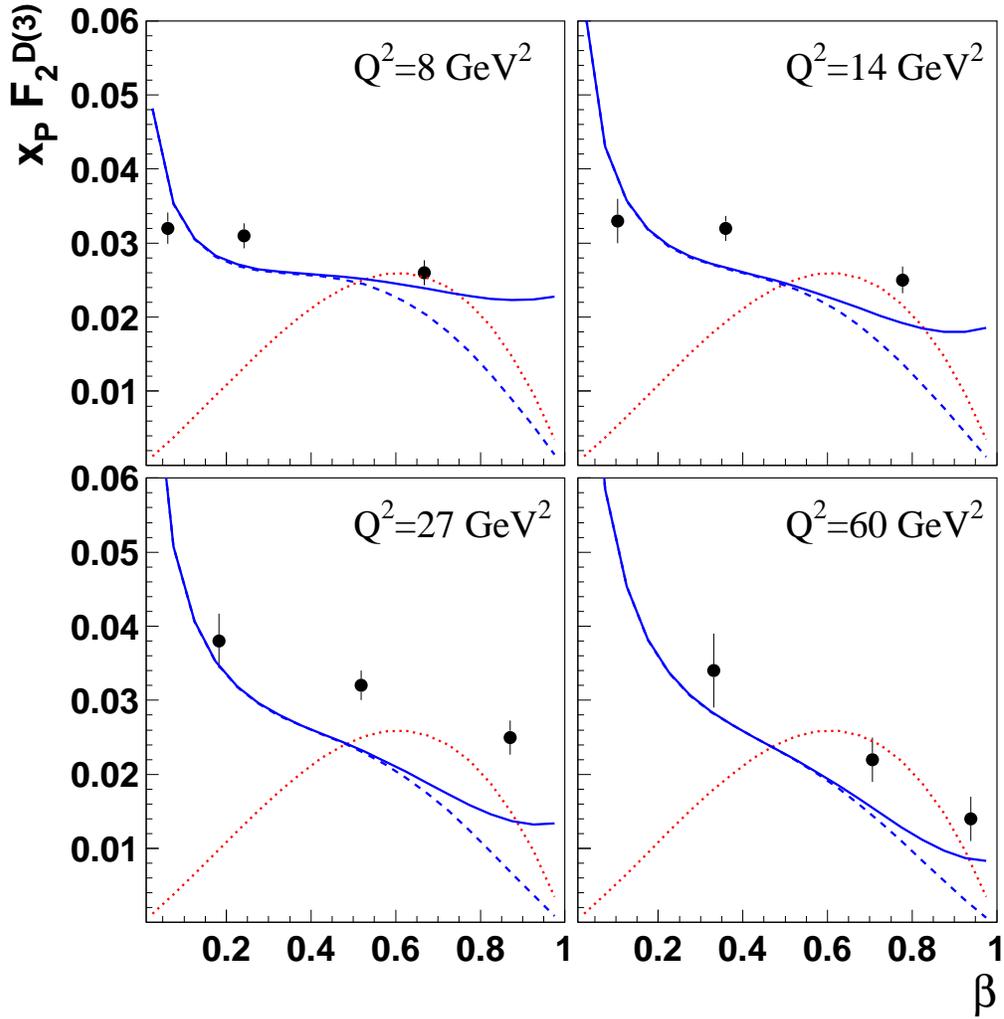,width=15cm}
               }
    \vspace*{-0.5cm}
\caption{The effect of the DGLAP evolution in $Q^2$. 
The dotted lines show the leading twist structure function at
the initial scale $Q_0^2=3~\mbox{\rm GeV}^2$. The dashed lines 
correspond to evolved structure function at the indicated
values of $Q^2$. The solid line is the sum of the evolved leading
twist 
contribution and twist-4 component, eq.~(\ref{eq:newanal}). 
The ZEUS data at $\xp=0.0042$ are shown.
} 
\label{fig6} 
\end{figure}

\end{document}